\documentclass[seceq]{ptptex}
\usepackage{wrapft}

\usepackage[dvips]{graphicx}

\title{
Possible Evolutionary Transition from Rapidly Rotating Neutron
Stars to Strange Stars Due to Spin-Down
}

\author{
Nobutoshi \textsc{Yasutake,}$^{1,}$\footnote{E-mail:
yasutake@gemini.rc.kyushu-u.ac.jp}
 Masa-aki \textsc{Hashimoto}$^{1,}$\footnote{E-mail:
hashi@gemini.rc.kyushu-u.ac.jp}\\
and Yoshiharu \textsc{Eriguchi}$^{2,}$\footnote{E-mail:
eriguchi@esa.c.u-tokyo.ac.jp}
}

\inst{
$^1$Department of Physics, Kyushu University, Fukuoka 810-8560, Japan\\
$^2$Department of Earth Science and Astronomy, Graduate School of
Arts and Sciences, University of Tokyo, Tokyo 153-8902,
Japan
}

\recdate{November 5, 2004      
}

\abst{
We present a scenario for the formation of strange stars due to spin-down of
{\it rapidly rotating} neutron stars left after supernova explosions.
In this scenario the rapid rotation plays a crucial role.
We assume that the total baryon mass is conserved but 
that
both total energy 
and angular momentum are lost due to emission of gravitational
waves and/or magnetic braking. Under this assumption,
we calculate the transition from rapidly rotating neutron stars to 
slowly rotating
strange stars. As a result, we find that a large amount of energy, 
$\sim 10^{53}$, ergs  is released. The liberated energy might become
a new energy source for the delayed explosion of a supernova. Furthermore, our
scenario suggests that supernovas associated with gamma-ray bursts are
feasible sources of observable gravitational
waves.

}

\begin{document}

\maketitle

\section{Introduction}

Since the possible existence of strange quark matter in nature was pointed
out \cite{ref:bod,ref:wit}, strange stars, that is, quark stars, 
have been examined as the most plausible sources of strange quark 
matter \cite{alo}. Concerning the origin and/or formation of strange stars 
\cite{ref:iid}, many researchers have investigated the relation between 
neutron stars and strange stars because it is natural to consider 
strange stars as compact objects left after supernova explosions, such as
the soft X-ray source RX J1856.5-3754 \cite{ref:koh03,ref:bom04,ref:web04}. 
As one other formation process, the transition process and the 
timescale from neutron stars to quark stars have been studied in detail 
from the point of view of pulsar spin-down due to strong magnetic fields \cite{har04}. 
If this kind of scenario is correct, there should be a process during which 
the maximum density of a neutron star increases to a larger value above 
which a strange star can be realized, i.e. the minimum density of a strange 
star, $\rho_c > 10^{15}\rm~g~cm^{-3}$ \cite{ref:wit}. This kind of process has 
already been investigated by considering accreting stars which shrink to
higher maximum energy densities, 
while there are large uncertainties in the physical process
involved~\cite{ref:bom00, ref:berez}.

The above described scenario could be related to supernova explosions
and gamma-ray bursts. In recent years, some gamma-ray bursts have been 
found to be related to supernova explosions, such as the connection between
SN1998bw and GRB980425 \cite{ref:gal}. The total energy of an 
event necessary to power sources at cosmological distances is on the order of a
few $10^{53}$ ergs for a 
spherical explosion
\cite{ref:kul}. We note that to generate a jet in supernova, 
rapid rotation and strong magnetic fields are thought to be required  
before the bounce~\cite{ref:kota04,ref:yama04}. Among many models 
\cite{ref:mesz}, the gravitational collapse of a super massive neutron star 
to a black hole was presented to explain both the energy source and the 
small baryon contamination \cite{ref:vie}. On the other
hand, the transition to a strange star has been suggested as the energy source 
of gamma-ray bursts \cite{ref:bom00,ref:chen,ref:wan,ref:xu,ref:ouy}. Huge 
amounts of energy could be liberated due to the phase transition from a neutron
star to a strange star within a small region, $\sim$ 10 km. This energy release 
leads to the creation of a fireball through the process $\gamma\gamma 
\rightleftharpoons e^{+}+e^-$. The transitions may occur on various time 
scales on the order of ms to a few years. This idea of a phase transition from a neutron
star to a strange star is attractive to avoid baryon contamination, because 
the baryonic mass ($\le 10^{-5} M_{\odot}$) of a thin crust 
$M_{\rm crust}$ in a strange star is
negligiblly small
\cite{ref:chen}. As a 
consequence, the resulting fireball can be accelerated to the relativistic regime 
with a Lorentz factor $\Gamma \sim 500$ if the energy release from the phase 
transition is of the order of $10^{52}$ ergs \cite{ref:wan}.

By contrast, if color superconductivity is realized at the surface of a quark 
star, the glueball decay causes a fireball \cite{ref:ouy}. They have total 
durations of 2 -- 81 s; $t_p\sim E_p^{-0.5}$ where $t_p$ is the peak duration 
and $E_p$ is the peak energy. This relation agrees with observation
\cite{ref:fen}. Furthermore, the decay of axion-like particles into $e^+e^-$
or photons leads to an ultra-relativistic plasma at distances $10^2$ -- $10^4$ 
km, which could be relevant to type Ib/Ic supernovae associated with weak 
gamma-ray bursts \cite{ref:bere}. We note that one-dimensional simulations of
the explosion from neutron stars to strange stars show that the explosions 
are too baryon rich and the energies too small to create a 
fireball\cite{ref:fw98}.

Moreover, observations of GRB990705 and GRB011211 that reveal the afterglow
emission lines could be interpreted as events after supernova explosions
with time intervals from hours to years \cite{ref:berez}. Concerning
these phenomena, Berezhiani et al. \cite{ref:berez} proposed a model of the
transition of a pure hadronic star to a quark star. The central density 
$\rho_c$ is assumed to increase due to spin-down or mass accretion,
where they used the mass-radius relation of Drago and Lavagno
\cite{ref:drago01}. The energy release was calculated from the difference 
between the gravitational mass of a metastable hadronic star and that of the 
final stable hybrid star (or quark star) with the same baryonic mass 
$M_{\rm B}$ \cite{ref:bom00}. Though the observation of the significant delay 
between supernova explosions and gamma-ray bursts has not been established 
\cite{ref:lui}, it is desirable to consider the transition from neutron stars 
to strange stars 
by taking into account 
elementary physics. On the other hand, it has been suggested that the emission 
of gravitational radiation and/or magnetic dipole radiation could cause 
the spin-down of compact stars \cite{ref:wan}.

However, we should note that most of these calculations/estimates concerning
the transition
from neutron stars to strange stars
have been studied using the spherical configurations; for example, the 
Tolman-Oppenheimer-Volkoff equation was solved to obtain the structures of 
compact stars \cite{ref:drago04}. In other words, the calculations have not 
included rotation, physical processes such as strong electric/magnetic fields 
on the surface of strange stars, and multi-dimensional effects of general 
relativity.

Although some investigations have taken into account some of the effects 
mentioned above, two-dimensional simulations of collapse-driven supernovae 
with updated neutrino transport fails to 
show the occurrence of 
explosions \cite{ref:janka}. Though the calculations are
very sensitive to the dynamical treatment of the energy transport between
the neutrino energy and the kinetic energy, it is shown that 
if only a small amount of energy is added to the kinetic energy, 
delayed explosions can appear. Furthermore, the artificial deformation 
of a progenitor has produced a shock 
that 
reached a radius of more than 600 km 
within 230 ms after a bounce. It is now believed that rotation strongly affects 
the development of shock propagation.

In order to make the situation clear and obtain a fireball in numerical
simulations, we need to carry out three-dimensional hydrodynamical
calculations in the framework of general relativity and to include 
physical processes such as the neutrino transport described by the Boltzmann 
transport equation. Since such exact simulations are impossible at present, 
we need
other approaches
to deepen our understanding of the evolution of 
compact stars. It is noted that the slow rotation approximation for rotating 
equilibrium used by Glendenning and Weber~\cite{ref:glen92} 
cannot be applied to configurations 
near the mass shedding limit
~\cite{ref:gour99}.

Evolutionary sequences of rapidly rotating strange stars in hydrostatic 
equilibrium were investigated for 
uniformly rotating configurations \cite{ref:gour99}. 
It was shown that the ratio of the rotational energy to the potential energy 
$T/|W|$ for rapidly rotating massive strange stars are very large, 
$T/|W| \sim 0.2$, which indicates the onset of triaxial instabilities due to 
gravitational radiation reactions. This suggests that strange stars could 
be sources of gravitational radiation. Numerical simulations of the 
nonlinear evolution of a nonaxisymmetric bar-mode perturbation of rapidly 
rotating stars find losses of more than 10\% of the initial angular 
momentum through gravitational waves over a time interval corresponding to 30 times 
the initial rotational period, $P_0$ \cite{ref:shiba04}. Though it depends
on the strength of the gravitational radiation reaction force, they suggest 
that the formed ellipsoid, which is assumed to be a proto-neutron star, will 
emit significant amounts of gravitational waves to dissipate the energy and the angular 
momentum after a long period. We note that they assumed a polytropic equation
of state, and $0.2 < T/|W| < 0.26$ in the initial model of a rotating star.
Recently,  the nucleation of quark matter in cores of rotating neutron stars 
has been studied by Harko et al. \cite{har04}, who 
assumed only 
a sequence of Maclaurin spheroids. The spin-down time scale was estimated to 
be $\sim 5$~s for an angular velocity $8000~\rm rad~s^{-1}$ and a magnetic 
field $\sim 10^{16}$~G.

In this paper, we present a new scenario for the transition from {\it a 
rapidly rotating neutron star} to {\it a slowly rotating strange star}
due to spin down through the emission of gravitational radiation and/or magnetic 
braking. Rapidly rotating compact stars are briefly discussed in \S 2 with an 
adopted equation of state. Our approach is qualitatively in advance to 
investigate the structures in the quasi-static evolution of the compact stars, 
because we need to obtain two-dimensional equilibrium configurations of 
{\it rapidly rotating} compact stars 
by fully taking into account 
general relativity up to the mass shedding
limits.
In \S 3, we show  that a significant amount of energy is released during the transition 
and predict the gravitational radiation accompanied by a delayed explosion 
of a supernova. In \S 4, we discuss the results of our scenario.

\section{Rapidly rotating compact stars}

Though the proto-neutron star left after a supernova explosion is very hot,
$T\sim50\rm~MeV$, it cools down to a cold neutron star ($T\sim 1~\rm MeV$) 
in some tens of seconds \cite{ref:prak}. Thus, during a time of a 
few hours after a supernova explosion, it is reasonable for a proto-neutron 
star to cool down to a cold neutron star. For a cold neutron star, we use a 
typical equation of state (EOS) derived by Wiringa, \cite{ref:wirin} which includes a 
three-body force. For strange stars, we adopt an EOS of strange matter 
based on the MIT bag model. Considering the uncertainties of the EOS for the 
strange matter, we adopt the simple EOS \cite{ref:gour99}
\begin{equation}
P=\frac13 (\rho - 4B), \ \ \
\rho = a n^{4/3} + B
 \ , \label{eq:eosb}
\end{equation}
where $\rho$ is the energy density, $n$ is the baryon number density in
units of $\rm fm^{-3}$, $a=952.4\rm~MeV~fm$, and $B$ is the bag constant in
$\rm MeV~fm^{-3}$. While we ignore both the strange quark mass and quark
interactions, the quantity $B$ is included to represent confinement effects.
The quantity $B$ is considered to be in the range $60 - 120\rm~MeV~fm^{-3}$. 
From the observation of RX J1856.5-3754 \cite{ref:drake},
Kohri et al. \cite{ref:koh03} derived an upper limit on the mass of a quark 
star for various sets of bag-model parameters. In this paper, we 
choose
three cases, $B=$~60, 90, and 120 $\rm ~MeV~fm^{-3}$. The corresponding maximum 
gravitational masses (proper radii) for spherical strange stars are 2.63 (10.7), 1.94 (8.75), 
and 1.56 $M_{\odot}$ (7.58 km), respectively. Therefore, the EOS (\ref{eq:eosb}) 
can cover the range of maximum masses that are obtained from more complex
EOSs of strange matter \cite{ref:koh03,ref:drago01,ref:drago04,ref:madb}.
It should be noted that there is no crust in a strange star expressed by the 
EOS (\ref{eq:eosb}), and therefore there is no baryonic contamination during the 
transition from a neutron star to a strange star. We note also that our 
results can be applied to a quark star of the type investigated by Ouyed and Sannino 
\cite{ref:ouy} as inner engines for gamma-ray bursts because they used almost
the same EOS (\ref{eq:eosb}), with 
$M_{\rm crust}\simeq5\times10^{-5}~M_{\odot}$.

With the same method used to investigate equilibrium configurations of
rapidly rotating stars \cite{ref:noza,ref:koma,ref:hashi}, we can obtain 
equilibrium sequences under the condition of stability for axisymmetric 
collapse as 
\begin{equation}
\left.\frac{\partial M}{ \partial \rho_{\rm max}} \right| _J > 0 \ , 
\label{eq:dmdro}
\end{equation}
where $J$ is the total angular momentum \cite{ref:haya98,ref:haya99}.
An equilibrium model can be obtained by computing the equilibrium
structure for a fixed $\rho_c$ and a given ratio $r_{p}/r_{e}$ of the polar
radius ($r_p$) to the equatorial radius ($r_e$). The stability condition for 
secular instability due to gravitational radiation emission is 
$T/|W| \gtrsim 0.14$. The value 0.14 is an upper limit, because 
instability for smaller wavelengths sets in for lower values of $T/|W|$.

The timescale $\tau_{\rm sec}$ of the growth of the secular instability due
to gravitational radiation can be estimated \cite{ref:fr75}. If we take the 
critical value for the onset of the instability, 
$\left(T/|W|\right)_{\rm c}=0.14$, we get $\tau_{\rm sec}$ for specific 
problems: $\tau_{\rm sec} \sim 6\times10^{-4}$ s for a gravitational mass 
$M\sim 1~M_{\odot}$, a proper radius $R\sim 10$ km, and $T/|W|\sim 0.21$.
Shibata and Karino~\cite{ref:shiba04} have obtained the growth time scale 
of the instability as $(3.5 - 31)\times P_0$ and estimated the evolution time scale 
of a proto-neutron star due to the gravitational radiation emission to be 
$\sim12~{\rm s}~(R/20~{\rm km})^4 (M/1.4 M_{\odot})^{-3}$. Thus, we can 
regard the evolution of a compact star born after supernova as
a sequence of equilibrium states for constant $M_B$, as studied for rotating proto-
neutron stars \cite{ref:vil}. A numerical simulation \cite{ref:shiba04}
shows that $J$ is dissipated  due to gravitational waves by $\sim 10 \%$ 
at the time $t=30~P_0$ for a neutron star of $\sim1.4 M_{\odot}$. For the rapidly 
rotating case, a neutron star formed after a supernova explosion will lose 
a significant amount of angular momentum in a rather short period 
$t < 0.03$ s. Therefore, before the transition from a neutron star to a strange
star, some amount of the total energy and angular momentum could
be lost. This would lead to an increase in $\rho_c$.

\begin{table}
\caption{Physical quantities for $B=~60~\rm MeV~fm^{-3}$ and $M_{\rm B}=2.6 
M_{\odot}$.}
\label{table:1}
\newcommand{\m}{\hphantom{$-$}}
\newcommand{\cc}[1]{\column{1}{c}{#1}}
\renewcommand{\tabcolsep}{0.5pc} 
\renewcommand{\arraystretch}{1.0} 
\begin{center}
\begin{tabular}{cccccccccl}
\hline \hline
Model      & $M$ &$n$&$r_p/r_e$&$r_e$&$J$ &$T/|W|$
&$\Delta E$ \\
\hline
 NS$^{*}$  & \m2.251 & $0.72$ & $0.575$ &14.5& $3.06$ & $0.309$ &
\\
 NS        & \m2.249 & $0.75$ & $0.668$ &12.9& $2.80$ & $0.117$ &
$0.036$ \\
 SS        & \m1.975 & $0.76$ & $0.892$ &11.7& $1.50$ & $0.036$ &
$ 5.1 $ \\
 SS'       & \m1.948 & $0.95$ & $ 1  $  &10.9& $ 0  $ & $  0  $ &
$ 0.31$ \\
\hline
\end{tabular}
\end{center}
\medskip
Here, NS$^*$ represents a neutron star of maximum rotation.
NS and SS describe the situations before and after the phase transition from a rotating 
neutron star to a strange star.
SS' represents a strange star without rotation. $\Delta E$ is the total energy
released in the transitions from NS$^*$ to NS, NS to SS, and SS to SS'.
The total angular momentum $J$ is in units of $\rm10^{49}~\rm 
g~cm^2~s^{-1}$, and $\Delta E$ is in units of $\rm 10^{53}~ergs$.
The maximum energy release, $\Delta E_{max}$ between 
NS$^*$ and SS' is 5.45 in these units. 
\end{table}

\begin{table}
\caption{Same as Table 1 but for $B=90~\rm MeV~fm^{-3}$ and $M_{\rm B}=1.7
M_{\odot}$. $\Delta E_{max}=1.92$ .}
\label{table:2}
\begin{center}
\begin{tabular}{ccccccccccl}
\hline \hline
Model      & $M$ &$n$&$r_p/r_e$&$r_e$&$J$ &$T/|W|$
&$\Delta E$ \\
\hline
 NS$^{*}$  & $1.547$ & $0.53$ & $0.575$ &15.7& $1.43$ & $0.111$ &
\\
 NS        & $1.533$ & $0.54$ & $0.706$ &13.4& $1.30$ & $0.087$ &
$0.25$ \\
 SS        & $1.462$ & $0.73$ & $0.797$ &9.90& $1.00$ & $0.074$ &
$ 1.3 $ \\
 SS'       & $1.436$ & $0.86$ & $ 1  $  &9.11& $ 0  $ & $  0  $ &
$ 0.47$ \\
\hline
\end{tabular}\\[2pt]
\end{center}
\end{table}

\begin{table}
\caption{Same as Table 1 but for $B=120\rm~MeV~fm^{-3}$ and $M_{\rm B}=1.5
M_{\odot}$. $\Delta E_{max}=0.641$ .}
\label{table:3}
\begin{center}
\begin{tabular}{ccccccccccl}
\hline \hline
Model      & $M$ &$n$&$r_p/r_e$&$r_e$&$J$ &$T/|W|$
&$\Delta E$ \\
\hline
 NS$^{*}$  & $1.379$ & $0.49$ & $0.581$ &15.8& $1.13$ & $0.105$ &
\\
 NS        & $1.375$ & $0.50$ & $0.695$ &13.7& $1.00$ & $0.086$ &
$0.071$ \\
 SS        & $1.366$ & $1.19$ & $0.935$ &8.11& $0.50$ & $0.026$ &
$0.16 $ \\
 SS'       & $1.343$ & $1.33$ & $ 1  $  &7.85& $ 0  $ & $  0  $ &
$0.41 $ \\
\hline
\end{tabular}\\[2pt]
\end{center}
\end{table}

\section{Results }

The transition density is unknown because of our lack of understanding of the basic 
physics concerning quark matter in a hadronic medium \cite{ref:bom04}.
Therefore, we assume that the transition due to the increase in the maximum 
density from a neutron star to a strange star occurs from an equilibrium state 
of a rapidly and uniformly rotating neutron star to another equilibrium
state of a slowly rotating strange star with i) the baryon mass remaining constant
and ii) the total energy and the angular momentum decreasing through 
gravitational radiation and/or magnetic dipole radiation.
This prescription is essentially the same as that used for the evolution of
an isolated rapidly rotating strange star that loses total energy and 
angular momentum \cite{ref:gon}.

\begin{figure}
\centering
\hspace*{25mm}
\includegraphics[bb=40 60 600 545, height=100mm, width=130mm]{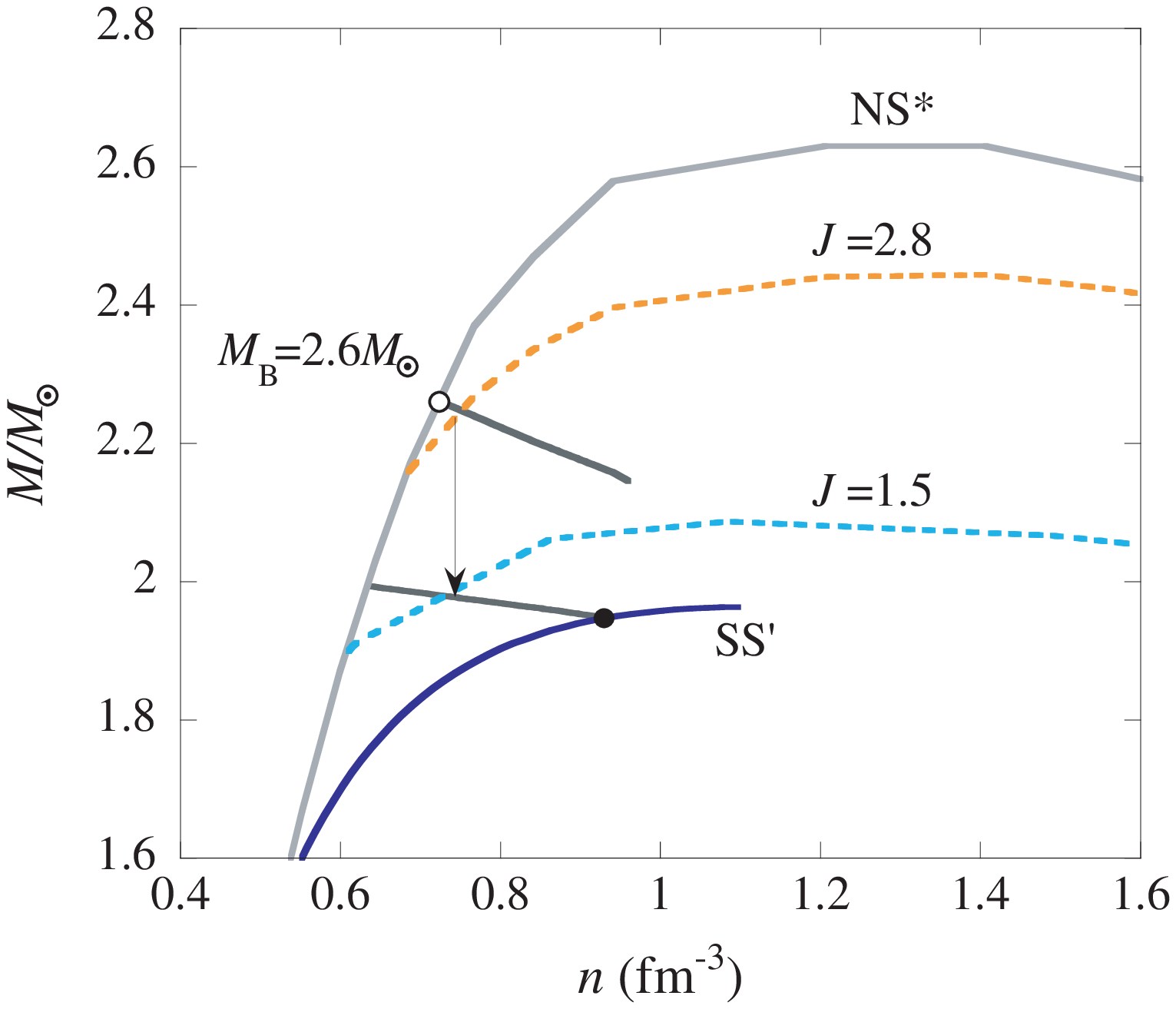}
\vspace{1.0cm}
\caption{An example of evolutionary transitions from NSs (upper thick and 
dashed curves)
to SSs (lower thick and dashed curves, for which $B=~60~\rm MeV~fm^{-3}$) with a fixed
baryonic mass of
$M_{\rm B} =2.6 M_{\odot}$ (two curves decreasing in the direction of increasing $n$). 
The total angular momentum $J$ (in units of
$10^{49}~\rm g~cm^2~s^{-1}$) is assumed to decrease due to gravitational
radiation during the transition starting from the $J=3.06$ state,
via $2.8$ and $1.5$ states, and finally reaching the $J=0$ state.
The arrow indicates the transition from NS to SS.
Although other transition paths are possible, the path must be between the 
points indicated by $\circ$ and $\bullet$.}
\label{fig:fig1}
\end{figure}

The results for the three values $B=$~60, 90, and 120 $\rm MeV~fm^{-3}$ are listed 
in Tables \ref{table:1} -- \ref{table:3} and in Figs \ref{fig:fig1}$\sim$\ref{fig:fig3}. 
We present an detailed scenario of the stellar transition 
based on the models of {\it rapidly rotating} compact stars. Instead of the method 
of Berezhiani et al., \cite{ref:berez} who used $M_{\rm B}$ to look for 
equilibrium sequences 
of 
spherical stars, we the use two parameters $J$ and $M_{\rm B}$ to find 
evolutionary sequences of uniformly rotating stars. Because there are many 
uncertainties in the transition mechanism, we present possible evolutionary 
paths.

Figure \ref{fig:fig1} depicts the evolutionary transition with $B=$~60 $\rm
MeV~fm^{-3}$ and $M_{\rm B}=2.6 M_{\odot}$ in the [gravitational mass 
$M$] -- [baryon number density $n$] plane. The two curves that decrease in 
the direction of increasing $n$ 
represent evolutionary tracks with constant 
$M_{\rm B}$. Our scenario is as follows. First, a neutron star with maximum 
rotation (NS$^*$) spins down due to gravitational radiation (or magnetic dipole 
radiation) emission ($J=3.1\longrightarrow2.8$, $\Delta E = 
3.6\times10^{51}$~ergs). Second, the NS changes into a slowly rotating SS 
($J=1.5$, $\Delta E = 5.1\times10^{53}$~ergs). Finally, the SS becomes a 
non-rotating SS' ($J=0$, $\Delta E = 3.1\times10^{52}$~ergs).
Though other evolutionary paths are possible, any path must lie between 
NS$^*$ and SS'. The total energy release, $\Delta E$, is calculated as the
difference between the gravitational masses
of the two states.
We note that the amounts of $\Delta E$ depend on $M_{\rm B}$.
We look for the largest value of $\Delta E$ by considering a 
large baryonic mass of $M_{\rm B} \geq 1.5 M_{\odot}$, as suggested by 
the evolutionary calculation of massive stars \cite{ref:hashi95}.
Physical quantities at the transition points are tabulated in 
Table \ref{table:1}.

The decrease in $\Delta E$ due to the angular momentum loss $\Delta J$ is 
approximately
\begin{equation}
\Delta E = \overline{\Omega} \Delta J ,
\end{equation}
where $\overline{\Omega}$ is the averaged angular velocity. Let us assume
that $J$ is lost through gravitational radiation. For the transition 
between neutron stars, if we adopt the value
$\Delta J \sim 10^{49}$ $\rm g~cm^2~s^{-1}$ and
$\overline{\Omega} \sim 10^3$ s$^{-1}$, then we have $\Delta E \sim 10^{52}$ ergs.
On the other hand, the relativistic energy loss rate due to the spin-down
of the magnetar of a proto-neutron star is estimated to be 
$\ge 10^{51}~\rm erg~s^{-1}$, and the emitted energy is 
$10^{51} - 10^{52}$~ergs on a time scale of 10--100 s \cite{ref:thom04}.
This energy release may not be sufficient to account for the gamma-ray burst 
energetics. By contrast, in the transition [neutron stars] $\longrightarrow$ 
[strange stars], an enormous amount of energy is released, as seen in Tables 
\ref{table:1} -- \ref{table:3}: $\Delta E = (0.16 - 5.1)\times10^{53}$ ergs. 
The nucleation time for $B=90\rm~MeV~fm^{-3}$ and $M=1.53$ $M_{\odot}$ is 
estimated to be 1 yr, \cite{ref:iid,ref:bom04} during which a metastable 
neutron star changes into a strange star. Harko et al.~\cite{har04} calculated 
the transition time to be 0.01~--~1 yr for a transition temperature of 
$T\sim0.01 - 1$~MeV. If the transition takes a long time, the energy release 
occurs in a clean environment without baryons. Since the energy release is 
radiated through gravitational radiation by at most 10\%, a significant amount 
of energy may be carried away by neutrinos. However, some amount of energy 
could be transferred to an explosion constituting of a supernova and/or produce a 
fireball that is contaminated by a small amount of baryons ($M_{\rm crust} 
\le 10^{-5} M_{\odot}$). The time scale of gravitational radiation from a 
realistic neutron star is estimated to be a few minutes, \cite{ref:shiba04}
and that of magnetic dipole radiation is 5--1000 s \cite{har04}.
Though these estimates are based on dimensional analysis, they are
consistent with the gamma-ray burst duration.

\section{Discussions}

We have shown that transitions from {\it a rapidly rotating neutron
star} to {\it a slowly rotating strange star} 
can be the energy source for a 
supernova explosion due to a large energy release of $\Delta E > 10^{53}$~ergs 
if $B \le 90$ $\rm MeV~fm^{-3}$, as seen in Tables \ref{table:1} and 
\ref{table:2}. While Bombaci et al. \cite{ref:bom04} obtained a liberated 
energy in the range $(0.5 - 1.7)\times10^{53}$ ergs, we obtain a maximum 
liberated energy of $5.1\times10^{53}$ ergs. We find that the transition 
energy between rotating stars exceeds that for non-rotating stars.
This is because the gravitational masses of a rotating stars exceed those 
of non-rotating stars.

Because no numerical simulations has 
shown a 
supernova explosion, a small amount of energy release is helpful to recover 
decaying shocks. Even if the available energy for a supernova is overestimated 
by a factor of 10 \cite{ref:fw98}, it is enough to explain hypernovae. Though 
the exact timescale concerning the angular momentum loss is unknown, the rapid 
rotation of compact stars would cause instabilities. However, we 
stress that this transition is accompanied by gravitational radiation and/or 
magnetic dipole radiation. In particular, the strength of the gravitational 
radiation determined from the rotation period/law may be related to the 
observed short gamma-ray bursts \cite{ref:hua}. Since gamma-ray bursts 
related to supernova explosions originate from jet-like explosions,
nonspherical configurations would result in 
gravitational radiation \cite{ref:kota04}.
Therefore, gamma-ray bursts associated with supernovae 
are feasible sources for the detection of gravitational waves.

Finally, we note that we have examined uniformly rotating configurations.
In fact, after a supernova explosion, a compact star would rotate differentially.
For $r_p /r_e =0.23$, a moderate differential rotation (rotation parameter
$A=1.0$ in the system studied by Komatsu et al. \cite{ref:koma}), and $\rho_c\sim 10^{15}~\rm 
g~cm^{-3}$, the gravitational mass is increased by a factor of 2.5 compared 
to the case of uniform rotation. Thus, the energy release obtained from 
differentially rotating neutron stars
is larger than the values of $\Delta E$ given in Tables \ref{table:1} -- \ref{table:3}.
On the other hand, differential rotations lead to significant deformations 
which are not only sources of gravitational radiation but are responsible 
for jet formation \cite{ref:yama04}. Therefore, even if pre-expelled 
supernova ejecta exist, a collimated jet could reach regions at large 
distances; a jet accelerated by rapid rotation can also accelerate
the pre-expelled material and leave an exit for the jet. It is suggested 
that a strong magnetic field related to the dipole radiation inherent in a 
rapid rotation creates a hole inside the supernova ejecta. The duration of 
gamma-ray bursts in the observer's frame due to external shocks is estimated 
to be 10 s for typical values of the initial energy, the mass loss rate, 
and the wind velocity with $\Gamma\sim100$ \cite{ref:mesz, ref:wan}. It is 
inferred that a newly born neutron star moves outward at a kick velocity of 
$\sim10^3$ km s$^{-1}$. If a phase transition occurs around the front of the 
supernova ejecta, the jet does not suffer from the contamination of the 
ejecta~\cite{ref:hua}. Considering the statics of high speed neutron stars 
(or strange stars), the detectable rate is estimated to be $10^{-4} - 10^{-6}$ 
per galaxy per year. Though the differential rotation law describing 
jet formation is not known, it is worthwhile to investigate that amount that 
differential rotation affects our transition scenario. Hydrodynamical 
calculations coupled with the dynamics of the phase transition from a
neutron star to a strange star are necessary to elucidate the effects 
presented in the present investigation.

\vskip 0.5cm
\noindent{\sl  Acknowledgements:}

{We thank  Shoichi Yamada for useful comments concerning the mechanism of
supernova explosion. 
We are grateful to the referees for comments and suggestions that helped us improve the paper. 
N. Yasutake thanks Shigeyuki Karino for helpful comments.}


\begin{figure}
\centering
\hspace*{20mm}
\includegraphics[bb=60 60 506 455, height=80mm, width=100mm]{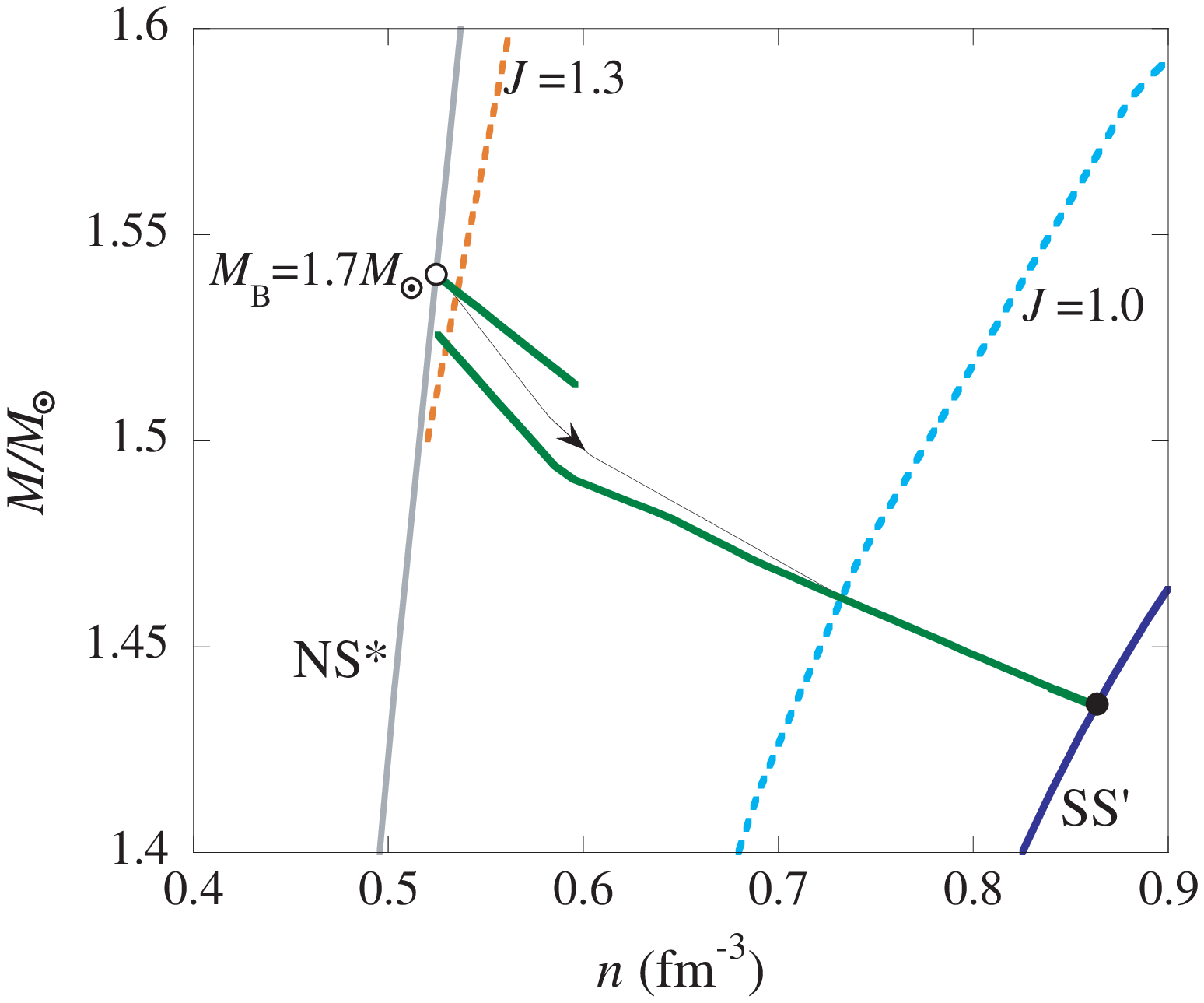}
\vspace{1.0cm}
\caption{Same as Fig. \ref{fig:fig1}, except with 
$B=~90~\rm MeV~fm^{-3}$, and $M_{\rm B} =1.7 M_{\odot}$, and
the transition from the $J=1.43$ state,
via $1.3$ and $1.0$ states, and reaching the $J=0$ state.}
\label{fig:fig2}
\end{figure}

\begin{figure}
\centering
\hspace*{20mm}
\includegraphics[bb=60 60 506 455, height=80mm, width=100mm]{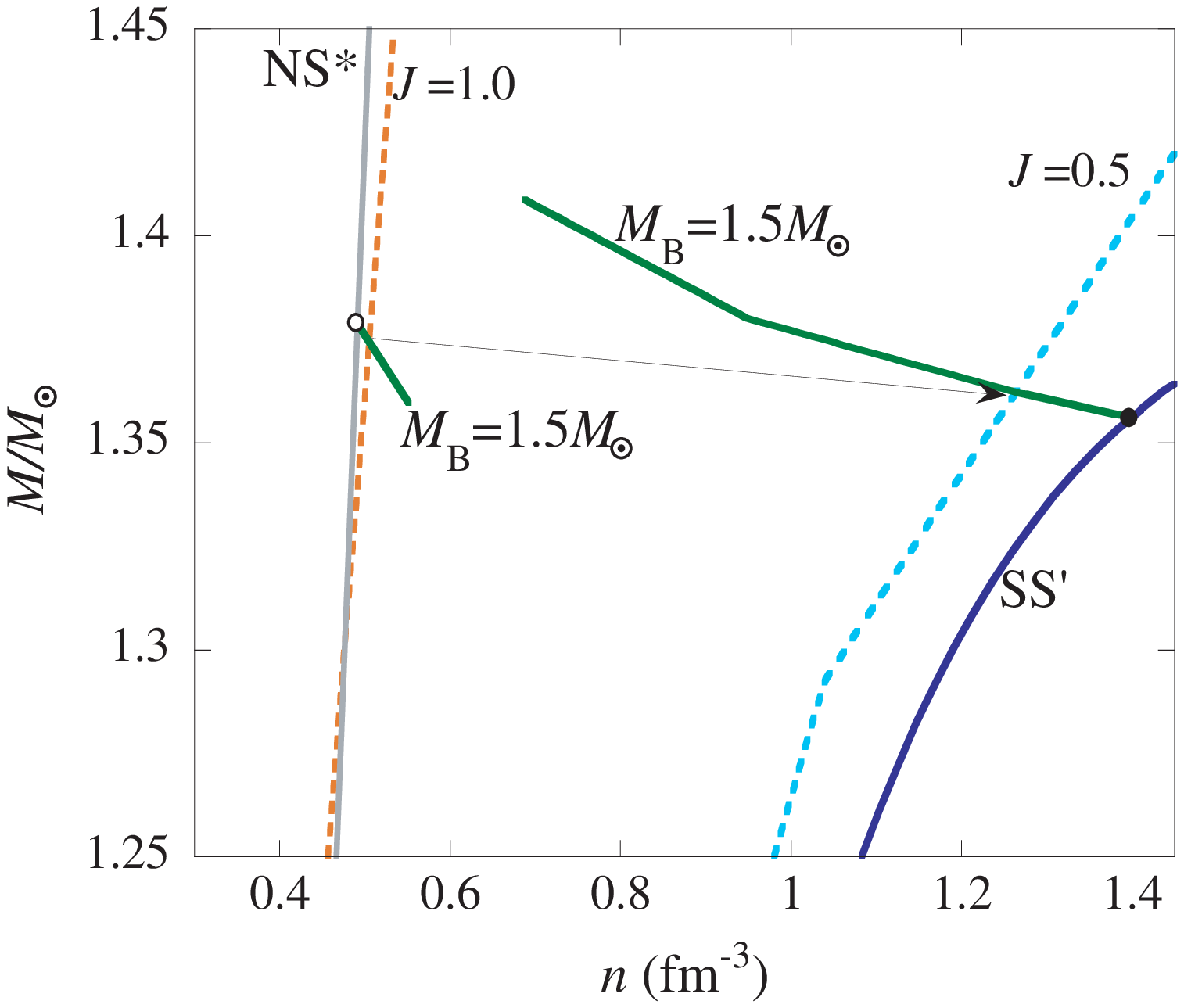}
\vspace{1.0cm}
\caption{Same as Fig. \ref{fig:fig1}, except with 
$B=~120~\rm MeV~fm^{-3}$ and $M_{\rm B} =1.5 M_{\odot}$, and
the transition from the $J=1.13$ state,
via $1.0$ and $0.5$ states and reaching the $J=0$ state.}
\label{fig:fig3}
\end{figure}

\end{document}